\documentstyle[12pt,a41,epsfig,portland,rotating,amstex,cite]{article}

\newcommand{\Li}{\mbox{Li}}

\newcommand{\Mvec}{\mbox{\rm\bf M}}

\newcommand{\beq}{\begin{equation}}
\newcommand{\eeq}{\end{equation}}
\newcommand{\bea}{\begin{eqnarray}}
\newcommand{\eea}{\end{eqnarray}}

\newcommand{\gsim}{\raisebox{-0.07cm}{$\, \stackrel{>}{{\scriptstyle
\sim}}\, $}}

\newcommand\GeV{\,\mbox{GeV}}

\newcounter{lin}

\graphicspath{{./},
	      {./figures/}}

\begin{document}
\begin{titlepage}

\begin{flushleft}
DESY 05--202 \hfill {\tt hep-ph/0608023} \\
SFB-CPP--06--36 \\
\end{flushleft}

\vspace{3cm}
\noindent
\begin{center}
{\Large\bf The Longitudinal Heavy Quark Structure Function 
\boldmath 
$F_{L}^{Q\overline{Q}}$}

\vspace*{2mm}
\noindent
{\Large \bf \boldmath in the Region $Q^2 \gg m^2$ at $O(\alpha_s^3)$}

\end{center}
\begin{center}

\vspace{2.0cm}
{\large J. Bl\"umlein$^a$, 
A. De Freitas$^a$\footnote{Alexander von Humboldt Fellow. Present address: 
Departamento de 
F\'{i}sica, Universidad Sim\'{o}n Bol\'{i}var, Apartado Postal 89000, 
Caracas 1080-A, Venezuela.},
 W.L. van Neerven$^b$ and
S. Klein$^a$}

\vspace{1.5cm}
{\it 
$^a$Deutsches Elektronen--Synchrotron, DESY}\\

\vspace{3mm}
{\it  Platanenallee 6, D--15738 Zeuthen, Germany}\\

\vspace{7mm}
{\it $^b$Instituut--Lorentz, Universiteit Leiden,}\\

\vspace{3mm}
{\it P.O. Box 9506, 2300 HA 
Leiden, The Netherlands}\\

\vspace{3cm}
\end{center}
\begin{abstract}
\noindent
The logarithmic and constant contributions to the Wilson coefficient of the 
longitudinal heavy quark structure function to $O(\alpha_s^3)$ are calculated 
using mass factorization techniques in Mellin space. The small $x$ behaviour
of the Wilson coefficient is determined. Numerical illustrations are 
presented.
\end{abstract}

\end{titlepage}

\newpage
\sloppy

\section{Introduction}
\label{sec:introduction}

\vspace{1mm}\noindent
Deeply inelastic electron--nucleon scattering at large momentum transfer
provides one of the cleanest possibilities to test the predictions of
Quantum Chromodynamics (QCD). In the case of pure photon exchange the 
structure functions $F_2(x,Q^2)$ and $F_L(x,Q^2)$ describe the scattering 
cross section. While the former structure function is well measured in a 
wide kinematic region \cite{PDG}, $F_L(x,Q^2)$ was mainly measured in 
fixed target 
experiments \cite{FTARG} and determined in the high $y$ region at HERA 
\cite{H1} using an extrapolation method. Future detailed measurements of the 
longitudinal structure function $F_L(x,Q^2)$ at HERA are still to be performed 
\cite{MK}. At leading order in the coupling constant the gluon distribution 
$g(x,Q^2)$ does not contribute to the structure function $F_2(x,Q^2)$ 
directly, but only to its derivative, which weakens the sensitivity. 
In the region of smaller values of $x$ the 
structure function $F_L(x,Q^2) = F_2(x,Q^2) - 2xF_1(x,Q^2)$, however, 
is  dominated by the gluon contribution. Therefore, this 
structure function may yield essential constraints on $g(x,Q^2)$.
In lowest order in the coupling constant ($\alpha_s^0$) and vanishing 
target--mass effects, the twist--2 contributions to the structure 
functions $F_2$ and $F_L$ obey the Callan--Gross \cite{CG} relation
\begin{eqnarray}
F_2(x,Q^2) = 2x F_1(x,Q^2), \hspace{1.5cm} F_L(x,Q^2) \equiv 0~. 
\end{eqnarray}
$F_L(x,Q^2)$ receives leading order contributions due to target mass 
effects \cite{GP}. The Callan--Gross relation is further broken by QCD 
corrections. The corresponding Wilson coefficients for massless quarks 
were calculated in leading  (LO) 
\cite{FLLO}, next-to-leading (NLO)  \cite{FLNLO,ZN,MV}, and 
next-to-next-to-leading order (NNLO) 
\cite{FLNNLO-mom,FLNNLO-xa,FLNNLO-xb}. Since the leading order coefficient 
functions are polynomial, scheme--invariant quantities one may 
construct a simple mapping of $F_L^{\rm LO}(x,Q^2)$ to 
$g^{\rm LO}(x,Q^2)$ taking the quark distributions from the $F_2$ 
measurement~\cite{DEVEN}.
The leading small $x$ terms for the coefficient functions of $F_L$ 
have been derived in \cite{CH} and agree with the known fixed order 
results (NLO, NNLO)
\cite{FLNLO,ZN,FLNNLO-xa,FLNNLO-xb}. The gluonic 
contribution to $F_L(x,Q^2)$ was calculated using the $k_\perp$ 
representation in leading order \cite{JBFL}, which turns out to be 
numerical very close to the NLO result \cite{FLNLO}. 
The numerical impact of the small $x$ resummation \cite{CH}
on $F_L$ was studied in \cite{BV}. Similar to the small $x$ resummation 
for the splitting functions, formally sub--leading terms lead to comparable 
but widely compensating effects, as seen comparing the magnitude of these 
terms for fixed orders in the coupling constant. This behaviour was 
later observed also in \cite{FLNNLO-xb}.  To draw firm conclusions 
on the effect of these resummations, several sub--leading series of terms 
have to be known.   Higher twist contributions 
to $F_L(x,Q^2)$, partly under model assumptions,  were considered in 
\cite{HTW}.  

Since the longitudinal structure functions $F_{L}(x,Q^2)$ contains 
rather 
large
heavy flavor contributions in the small $x$ region \cite{BR}, a consistent 
analysis has to account for these effects, which were calculated in 
leading \cite{FLQQLO} and next-to-leading order 
\cite{FLQQNLO1,FLQQNLO2}.\footnote{Fast Mellin--space expressions for 
these Wilson coefficients were given in \cite{SAJB}.} The NLO  corrections 
\cite{FLQQNLO1} could not be performed in analytic form completely. This 
is also expected for even higher orders, due to the complexity of the 
phase space integrals. However, complete analytic results may be derived 
in the asymptotic region $Q^2 \gg m^2$ calculating all contributions 
but the power suppressed terms $(m^2/Q^2)^k$,~~\cite{BUZA1, 
BUZA2}.\footnote{For related work for other processes, see  
\cite{BUZA3}.}

In the present paper we use the method of Ref.~\cite{BUZA1} to derive
the heavy quark Wilson coefficients for $F_L^{Q\overline{Q}}(x,Q^2)$ to 
$O(\alpha_s^3)$ in the region $Q^2 \gg m^2$. In Section~2 we give a brief 
outline of the method. The Wilson coefficients are derived in 
Section~3. Their small $x$ behaviour is discussed in Section~4. In 
Section~5 numerical are presented and  Section~6 contains 
the conclusions. Some useful relations are summarized in 
the Appendix.
\section{\boldmath The Method}
\label{sec:METH}

\vspace{1mm}\noindent
In the twist--2 approximation the nucleon structure functions $F_i(x,Q^2)$
are described as Mellin convolutions between the parton densities 
$f_j(x,\mu^2)$ and the Wilson coefficients $C_i^j(x,Q^2/\mu^2)$
\begin{eqnarray}
\label{STR}
F_i(x,Q^2) = \sum_j C_i^j\left(x,\frac{Q^2}{\mu^2}\right) \otimes 
f_j(x,\mu^2)
\end{eqnarray}
to all orders in perturbation theory due to the factorization theorem. 
Here $\mu^2$ denotes the factorization scale and
the Mellin convolution is given by the integral
\begin{eqnarray}
[A \otimes B](x) = \int_0^1 dx_1 \int_0^1 dx_2~~ \delta(x - x_1 x_2) 
~A(x_1) B(x_2)~.
\end{eqnarray}
Since the distributions $f_j$ refer to {\sf massless} partons, the heavy 
flavor effects are contained in the Wilson coefficients only.
We are interested in the massive contributions in the region $Q^2 \gg 
m^2$. These are the non--power corrections in $m^2/Q^2$, i.e. all 
logarithmic contributions and the constant term. We apply the collinear 
parton model, i.e. the parton 4--momentum is $p = z P$, with  
$P$ the nucleon momentum. The massive Wilson coefficients itself
can be viewed as a quasi cross section in $p V^*$--scattering, where $V^*$ 
denotes the exchanged virtual vector boson. In the limit $Q^2 \gg m^2$ the 
massive Wilson coefficients $H^{\rm S,NS}_{2,L,i}(Q^2/m^2,m^2/\mu^2,x)$,
likewise the case for the structure functions (\ref{STR}), factorize 
\begin{eqnarray}
\label{HFAC}
H_{2,L,i}^{\rm S, NS}\left(\frac{Q^2}{m^2}, \frac{m^2}{\mu^2},x\right) = 
C_{2,L,k}^{S,NS} \left(\frac{Q^2}{\mu^2},x\right) 
\otimes	A_{k,i}^{\rm S, NS} \left(\frac{m^2}{\mu^2},x\right) 
\end{eqnarray}
into Wilson coefficients $C_{L,k}^{S,NS} \left(Q^2/\mu^2,x\right)$ 
accounting for light flavors only and  massive operator matrix 
elements $A_{k,i}^{\rm S, NS} \left(m^2/\mu^2,x\right)$. The 
latter take a similar role as the parton densities in (\ref{STR}), but are
perturbatively calculable. The factorization (\ref{HFAC}) is a 
consequence of the renormalization group and the fact that we restrict the
investigation to non--power corrections to
$H_{L,i}^{\rm S, NS}\left(Q^2/m^2,m^2/\mu^2,x\right)$. 
The operator matrix elements $A_{k,i}^{\rm S, NS}$ obey the expansion
\begin{eqnarray}
\label{op1}
A_{k,i}^{\rm S, NS} \left(\frac{m^2}{\mu^2}\right) = \langle 
i|O_k^{\rm S, NS}|i\rangle
= \delta_{k,i} + \sum_{l=1}^{\infty} a_s^l A_{k,i}^{{\rm S, 
NS},(l)},~~~~i=q,g 
\end{eqnarray}
of the twist--2 quark singlet and non--singlet  operators $O_k^{\rm S, 
NS}$ between {\sf partonic} states 
$|i\rangle$, 
which are related by collinear factorization to the initial--state nucleon 
states $|N\rangle$ and $a_s = \alpha_s(\mu^2)/(4\pi)$ denotes the strong 
coupling constant. The operator matrix elements are process--independent 
quantities. 
The process dependence of $H_{L,i}^{\rm S, NS}$ 
is described by the associated coefficient functions 
\begin{eqnarray}
\label{co1}
{C}_{L,k} \left(\frac{Q^2}{\mu^2}\right) = 
\sum_{l=l_0}^{\infty} a_s^l C_{L,k}^{(l)}
\left(\frac{Q^2}{\mu^2}\right), \hspace{7mm} k= {\rm NS, S}, g~.
\end{eqnarray}
The $\overline{\rm MS}$ 
coefficient functions, in the massless limit, corresponding to the heavy 
quarks only, are denoted by
\begin{eqnarray}
\widehat{C}_{L,k}\left(\frac{Q^2}{\mu^2}\right) = 
C_{L,k}\left(\frac{Q^2}{\mu^2},N_L+N_H\right)  
- C_{L,k}\left(\frac{Q^2}{\mu^2},N_L\right)~, 
\end{eqnarray}
where $N_H, N_L$ are the number of heavy and light flavors, 
respectively. In the following we will consider the case of a single heavy 
quark, i.e. $N_H = 1$. The formalism is easily generalized to more than 
one heavy quark species.
The heavy flavor Wilson coefficient is obtained as the expansion of 
the product of (\ref{op1},\ref{co1}) to the respective order in $a_s$. 
\section{\boldmath The Wilson Coefficients in the Region $Q^2 \gg M^2$}
\label{sec:WILSON}

\vspace{1mm}\noindent
In the limit of vanishing nucleon mass 
effects, cf.~\cite{GP}, the longitudinal structure function emerges only 
at $O(a_s)$ due to the Callan--Gross relation \cite{CG}. The leading order 
contribution is purely gluonic  \cite{FLLO}. At $O(a_s^2)$ 
$F_L^{Q\overline{Q}}(x,Q^2)$ receives also quarkonic contributions.

To $O(a_s^3)$ the heavy quark Wilson coefficients $H_{L,g(q)}^{\rm 
S, PS, NS}$ read~:
\begin{eqnarray}
\label{eq1a}
H_{L,g}^{\rm S}\left(\frac{Q^2}{m^2}, \frac{m^2}{\mu^2}\right)
&=& a_s    \widehat{C}^{(1)}_{L,g}\left(\frac{Q^2}{\mu^2}\right)
 +  a_s^2  \left[ 
A_{Q,g}^{(1)}\left(\frac{\mu^2}{m^2}\right) \otimes
                  C^{(1)}_{L,q}\left(\frac{Q^2}{\mu^2}\right)
 +                \widehat{C}^{(2)}_{L,g}\left(\frac{Q^2}{\mu^2}\right)\right] 
\nonumber\\
&+& a_s^3 \left[
A_{Q,g}^{(2)}\left(\frac{\mu^2}{m^2}\right) \otimes
                  C^{(1)}_{L,q}\left(\frac{Q^2}{\mu^2}\right)
+ A_{Q,g}^{(1)}\left(\frac{\mu^2}{m^2}\right) \otimes
                  C^{(2)}_{L,q}\left(\frac{Q^2}{\mu^2}\right)
 +                
\widehat{C}^{(3)}_{L,g}\left(\frac{Q^2}{\mu^2}\right)\right] 
\nonumber\\ \\
\label{eq1b}
H_{L,q}^{\rm PS}\left(\frac{Q^2}{m^2}, \frac{m^2}{\mu^2}\right)
&=& a_s^2  \widehat{C}^{{\rm PS},(2)}_{L,q}\left(\frac{Q^2}{\mu^2}\right)
 +  a_s^3  \left[ 
A_{Qq}^{{\rm PS},(2)}\left(\frac{\mu^2}{m^2}\right) \otimes
                  C^{(1)}_{L,q}\left(\frac{Q^2}{\mu^2}\right)
 +\widehat{C}^{{\rm PS},(3)}_{L,q}\left(\frac{Q^2}{\mu^2}\right)\right] 
\\
H_{L,q}^{\rm NS}\left(\frac{Q^2}{m^2}, \frac{m^2}{\mu^2}\right)
&=& a_s^2    \widehat{C}^{{\rm NS},(2)}_{L,q}\left(\frac{Q^2}{\mu^2}\right)
 +  a_s^3  \left[ 
A_{qq,Q}^{{\rm NS},(2)}\left(\frac{\mu^2}{m^2}\right) \otimes
                  C^{(1)}_{L,q}\left(\frac{Q^2}{\mu^2}\right)
 +                
\widehat{C}^{{\rm NS},(3)}_{L,q}\left(\frac{Q^2}{\mu^2}\right)\right]~, 
\end{eqnarray}
where
\begin{eqnarray}
{C}_{L,q}^{(2)} = C_{L,q}^{\rm NS} + C_{L,q}^{\rm PS}
\end{eqnarray}
and
\begin{eqnarray}
{H}_{L,q}^{\rm S} = {H}_{L,q}^{\rm NS} + {H}_{L,q}^{\rm PS}~. 
\end{eqnarray}
$C_{L,i}^{(k)}(Q^2/\mu^2)$ are 
the scale dependent Wilson coefficients in the $\overline{\rm MS}$ scheme with  
$C_{L,i}^{(k)}(Q^2/\mu^2) =  c_{L,i}^{(k)}$ for $Q^2 = \mu^2$ given 
in~\cite{FLLO,FLNLO,ZN,MV,FLNNLO-mom,FLNNLO-xa,FLNNLO-xb}. 

The operator matrix elements were derived in \cite{BUZA1} and read~:
\begin{eqnarray}
\label{eq1c}
A_{Qg}^{(1)} &=& - \frac{1}{2} \widehat{P}_{qg}^{(0)} 
\ln\left(\frac{m^2}{\mu^2}\right) + a_{Qg}^{(1)}\\
\label{eq1d}
A_{Qg}^{(2)} &=& \frac{1}{8}\left\{ \widehat{P}_{qg}^{(0)} \otimes 
\left[P_{qq}^{(0)} - P_{gg}^{(0)} + 2 \beta_0\right]\right\} \ln^2\left(
\frac{m^2}{\mu^2}\right)\nonumber\\
& & - \frac{1}{2} \left\{\widehat{P}_{qg}^{(1)} + a_{Qg}^{(1)} 
\left[P_{qq}^{(0)}
-P_{gg}^{(0)} + 2 \beta_0\right] \right\} \ln 
\left(\frac{m^2}{\mu^2}\right) \nonumber \\
& & + \overline{a}_{Qg}^{(1)} \left[P_{qq}^{(0)}
-P_{gg}^{(0)} + 2 \beta_0\right]  + a_{Qg}^{(2)}
\\
\label{eq1e}
A_{Qq}^{{\rm PS},(2)} &=& - \frac{1}{8} \widehat{P}_{qg}^{(0)} \otimes 
P_{gq}^{(0)}
\ln^2\left(\frac{m^2}{\mu^2} \right) - \frac{1}{2} \left[
\widehat{P}_{qq}^{{\rm PS},(1)} - a_{Qg}^{(1)} P_{gq}^{(0)} \right] 
\ln\left(\frac{m^2}{\mu^2}\right) \nonumber\\
& & + a_{Qq}^{{\rm PS},(2)} 
+ \overline{a}^{(1)}_{Qg} \otimes P_{gq}^{(0)} 
\\
A_{qq,Q}^{{\rm NS},(2)} &=& - \frac{\beta_{0,Q}}{4} P_{qq}^{(0)} \ln^2
\left(\frac{m^2}{\mu^2}\right) - \frac{1}{2} \widehat{P}_{qq}^{{\rm 
NS},(1)} \ln
\left(\frac{m^2}{\mu^2}\right) + a_{qq,Q}^{{\rm NS},(2)} 
+\frac{1}{4} \beta_{0,Q} \zeta_2 P_{qq}^{0}~, 
\end{eqnarray}
with
\begin{eqnarray}
\widehat{f} = {f}(N_F+1) - f(N_F)~.
\end{eqnarray}
For later fast numerical representations we express the above 
functions $f_i(x)$ in Mellin space, 
\begin{eqnarray}
\Mvec[f_i(x)](N) = \int_0^1 dx~x^{N-1}~ f_i(x)
\end{eqnarray}
at (even) integers $N$ and arrange for analytic continuation to complex 
values of $N$ starting from these values.

The splitting functions are
\begin{eqnarray}
P_{qq}^{(0)}(N) &=& 4 C_F\left[-2S_1(N-1) + \frac{(N-1) (3N+2)}{2 N 
(N+1)}\right]\\
P_{qg}^{(0)}(N) &=& 8 T_R N_F \frac{N^2 + N + 2}{N (N+1) (N+2)} 
\\
\label{eqPgg0}
P_{gg}^{(0)}(N) &=& 8 C_A\left[-S_1(N-1) - \frac{N^3 - 3 N -4}
{(N-1) N (N+1) (N+2)}\right] + 2 \beta_0
\\
\label{eqPgq0}
P_{gq}^{(0)}(N) &=& 4 C_F \frac{N^2 + N +2}{(N-1) N (N+1)}
\\
\widehat{P}_{qq}^{{\rm PS},(1)}(N) &=& 16 C_F T_R  \frac{5 N^5 + 32 N^4
                           + 49 N^3 +38 N^2 +28 N + 8}{(N-1) N^3 (N+1)^3 
(N+2)^2}
\\
P_{qq,Q}^{{\rm NS},(1)}(N) = \widehat{P}_{qq}^{{\rm NS},(1)} &=& C_F T_R 
\left\{\frac{160}{9} S_1(N-1) - 
\frac{32}{3} S_2(N-1)\right.\nonumber\\
&& \hspace{12mm} \left. -\frac{4}{9}\frac{(N-1)(3N+2)(N^2-11 N -6)}{N^2 
(N+1)^2}\right\}\\
\widehat{P}_{qg}^{(1)}(N) &=& 8 C_F T_R \left\{
                    2 \frac{N^2 + N + 2}{N (N+1) (N+2)} \left[S_1^2(N)
                    - S_2(N)\right] - \frac{4}{N^2} S_1(N) 
\right. \nonumber\\
& & \left. \hspace{1.3cm} + \frac{5 N^6 + 15 N^5 + 36 N^4 + 51 N^3 +25 N^2 
+ 8 N +4}{N^3 (N+1)^3 (N+2)}\right\}  \nonumber\\
& &+  16 C_A T_R \left\{
-\frac{N^2+N+2}{N(N+1)(N+2)} \left[S_1^2(N) 
+ S_2(N) - \zeta_2 - 2 \beta'(N+1) \right]
\right. \nonumber\\ 
& & \hspace{1.3cm} \left. + 4 \frac{2N +3}{(N+1)^2 (N+2)^2} S_1(N)  
 + \frac{P_1(N)}{(N-1) N^3 (N+1)^3 (N+2)^3}\right\}~, 
\nonumber\\
\end{eqnarray}
where
\begin{eqnarray}
P_1(N) &=& N^9 +6 N^8 + 15 N^7 + 25 N^6 + 36 N^5 + 85 N^4 + 128 N^3 
+ 104 N^2  \nonumber\\ & &
+ 64 N +16~.
\end{eqnarray}
The expansion coefficient of the $\beta$--function for the case of 
light and heavy $(Q)$ flavors read
\begin{eqnarray}
\beta_0     &=& \frac{11}{3} C_A - \frac{4}{3} T_R N_f~,\\
\beta_{0,Q} &=& -\frac{4}{3} T_R~.
\end{eqnarray}

The Mellin transforms lead to harmonic sums. Their analytic continuation 
for single harmonic sums is given by
\begin{eqnarray}
S_1(N-1) &=& \psi(N) + \gamma_E\\
S_k(N-1) &=& \frac{(-1)^{k-1}}{(k-1)!} \psi^{(k-1)}(N) + \zeta_k,~~~~~k 
\geq 
2~\\
S_{-1}(N-1) &=& (-1)^{N-1} \beta(N) - \ln(2) \\ 
S_{-k}(N-1) &=& (-1)^{k+N} \beta^{(k)}(N) - 
\left(1-\frac{1}{2^{k-1}}\right) \zeta_k,~~~~~k \geq 2~.  
\end{eqnarray}
Here $\psi(z) = d \ln[\Gamma(z)]/dz$, $\gamma_E$ denotes the 
Mascheroni--Euler number,  $\zeta_k$ the values of 
Riemann's $\zeta$--function for integer $k \geq 2$ and 
\begin{eqnarray}
\beta(z) = \frac{1}{2} \left[ \psi\left(\frac{1+z}{2}\right) - 
\psi\left(\frac{z}{2}\right)\right]~.
\end{eqnarray}
Multiply nested harmonic sums are reduced to Mellin transforms of basic 
functions \cite{BK1,ALGEBRA,JBVR} for which the analytic 
continuation to complex values of $N$ is performed \cite{ANCONT,JBSM}.

The functions emerging in the scale independent contributions of the 
operator matrix elements are
\begin{eqnarray}
a_{Qg}^{(1)}(N) &=& 0\\
\overline{a}_{Qg}^{(1)}(N) &=& -\frac{1}{8} \zeta_2 
\widehat{P}_{qg}^{(0)}(N)
%
\end{eqnarray}
\begin{eqnarray}
\label{eqaQg2}
a_{Qg}^{(2)}(N) &=& 4 C_F T_R \Biggl\{
                   \frac {{N}^{2}+N+2}{N \left( N+1 \right)  \left( N+2 \right) }
                   \Biggl[
                   - \frac{1}{3} S_1^3(N-1) + \frac{4}{3} S_3(N-1) 
\nonumber\\ & & \hspace{1.5cm}
                   - S_1(N-1) S_2(N-1) - 2 \zeta_2 S_1(N-1) \Biggr]
+ \frac{2}{N(N+1)} S_1^2(N-1) 
\nonumber\\ & & \hspace{1.5cm}
+  \frac{N^{4} +16\,{N}^{3} +15\,{N}^{2}-8\,N-4}
        {N^2 \left( N+1 \right)^{2} \left( N+2 \right) } S_2(N-1)
\nonumber\\ & & \hspace{1.5cm}
+\frac { 3\,{N}^{4}+2\,{N}^{3}+3\,{N}^{2}-4\,N-4}
      {2 N^2 \left( N+1 \right) ^{2} \left( N+2 \right) } \zeta_2
\nonumber\\ & & \hspace{1.5cm}
+\frac {N^4-N^3-16 N^2 + 2N +4}
       {N^2 \left( N+1 \right) ^{2}\left( N+2 \right)} S_1(N-1)
+ \frac {P_2(N)}
{2 N^4 \left( N+1 \right) ^{4} \left( N+2 \right) }\Biggr\}
\nonumber\\
& & +4 C_A T_R\Biggl\{ \frac{N^2+N+2}{N(N+1)(N+2)} 
\Biggl[ 4 \Mvec\left[\frac{\Li_2(x)}{1+x}\right](N) +\frac{1}{3} 
S_1^3(N) 
+
3 S_2(N) S_1(N) \nonumber\\ 
& & \hspace{1.5cm} 
+ \frac{8}{3} S_3(N) 
+ \beta''(N+1) - 4 \beta'(N+1) S_1(N) - 4 \beta(N+1) 
\zeta_2 
+\zeta_3\Biggr]\nonumber\\
& & \hspace{1.5cm}
-\,{\frac {{N}^{3}+8\,{N}^{2}+11\,N+2}{N \left( N+1 \right) ^{2} \left( N+2
 \right) ^{2}}} S_1^2(N)
-2\,{\frac {N^4 - 2 N^3 + 5 N^2+ 2 N + 2} 
{ \left( N-1 \right)  N^2 \left( N+1 \right) ^{2} \left( N+2 
\right) }} 
\zeta_2\nonumber\\
& & \hspace{1.5cm}
-\, \frac {7 N^5 + 21 N^4 + 13 N^3 + 21 N^2 +18 N +16} 
           { (N-1) N^2 \left( N+1 \right) ^{2} \left( N+2 \right) ^{2}} 
           S_2(N) 
\nonumber
\end{eqnarray}\begin{eqnarray}
& & \hspace{1.5cm}
-\,{\frac {{N}^{6}
 +8\,{N}^{5}
+23\,{N}^{4}
+54\,{N}^{3}
+94\,{N}^{2}
 +72\,N
  +8}{ N \left( N+1 \right) ^
{3} \left( N+2 \right) ^{3} }} S_1(N)
\nonumber\\ 
& & \hspace{1.5cm}
-4 \, \frac { \left( {N}^{2} - N -4 \right)}
            { \left( N+1 \right) ^{2} \left( N+2 \right) ^{2}} \beta'(N+1)
+ \frac{P_3(N)}{(N-1) N^4 (N+1)^4 (N+2)^4}\Biggr\}
\end{eqnarray}\begin{eqnarray}
\label{eqaQq2}
a_{Qq}^{{\rm PS},(2)}(N) &=& C_F T_R \left\{- 8 \frac{N^4 + 2 N^3 + 5 N^2 
+4 N + 4}{(N-1) N^2 (N+1)^2 (N+2)} S_2(N-1) 
\right. \nonumber\\ & & \hspace{12mm}
-4\frac{(N^2 + N + 2)^2}{(N-1)   
N^2 (N+1)^2 (N+2)} \zeta_2 
\left. + \frac{4~P_4(N)}{(N-1) N^4 (N+1)^4 (N+2)^3} \right\} \\
a_{qq,Q}^{{\rm NS},(2)}(N) &=& C_F T_R \Biggl\{-\left(\frac{224}{27} + 
\frac{8}{3} 
\zeta_2 \right) S_1(N-1) + \frac{40}{9} S_2(N-1) - \frac{8}{3} S_3(N-1) 
\nonumber\\ 
&& 
\hspace{12mm} + \frac{2 (3N+2)(N-1)}{3N(N+1)} \zeta_2
\nonumber\\ 
&& \hspace{12mm}
+ \frac{(N-1)(219 N^5 +428 N^4 +517 N^3 + 512 N^2 +312 N + 72)}
{54 N^3 (N+1)^3} \Biggr\}~, 
\end{eqnarray}
where
\begin{eqnarray}
P_2(N) &=& 12 N^{8}
             +54 N^{7} 
            +136 N^{6}
            +218 N^{5}
            +221 N^{4}
            +110 N^{3}
              -3 N^{2}
             -24 N
              -4
\\
P_3(N) 
    &=& 2\,{N}^{12}
      +20\,{N}^{11}
      +86\,{N}^{10}
     +192\,{N}^{9}
     +199\,{N}^{8}
          -{N}^{7}
     -297\,{N}^{6}
     -495\,{N}^{5}
     -514\,{N}^{4} \nonumber\\ & &
     -488\,{N}^{3}
     -416\,{N}^{2}
      -176\,N
       -32
\\
P_4(N) &=& N^{10} + 8 N^9 + 27 N^8 +33 N^7 - 71 N^6 - 275 
N^5 - 403 N^4 - 448 N^3\nonumber\\ & & - 408 N^2 - 208 N - 48~.
\end{eqnarray}
Note that in the above expressions the equality 
\begin{eqnarray}
\Mvec\left[\frac{\Li_2(x)}{1+x}\right](N)  - \zeta_2 \beta(N)
= (-1)^{N} \left[S_{-2,1}(N-1)  + \frac{5}{8} \zeta_3\right]
\end{eqnarray}
can be applied. Therefore, the  operator matrix 
elements depend on one non-trivial basic function \cite{JBVR,JB05} only 
and no sum with index $\{-1\}$ contributes. Concerning the 
non--trivial harmonic sums emerging the operator matrix elements are 
of the same complexity as the 2--loop anomalous dimensions.   

The longitudinal structure function $F_L(x,Q^2)$ consists of the light and 
heavy flavor contributions
\begin{eqnarray}
\label{eqFLa}
F_L(x,Q^2) &=& F_L^{\rm light}(x,Q^2) + F_L^{\rm heavy}(x,Q^2)
\nonumber\\
           &=& C_L^{\rm NS}\left(x,a_s,\frac{Q^2}{\mu^2}\right) \otimes 
               q_{\rm NS}(x, \mu^2)
            +  C_L^{\rm S}\left(x,a_s,\frac{Q^2}{\mu^2}\right) \otimes 
               q_{\rm S}(x, \mu^2) \nonumber\\ & &
            +  C_L^{g}\left(x,a_s,\frac{Q^2}{\mu^2}\right) \otimes 
               g(x, \mu^2)~.
\end{eqnarray}
We choose $Q^2 = \mu^2$ as uniform factorization scale.
The Wilson coefficients in (\ref{eqFLa}) are of the form
\begin{eqnarray}
C_L^{(i)}\left(x, a_s, \frac{Q^2}{\mu^2}\right) = 
C_L^{(i),{\rm light}}(x, a_s)  + 
H_L^{(i)}\left(x, a_s, \frac{Q^2}{m^2}\right),~~~~~i = {\rm S, NS},~g~. 
\end{eqnarray}
The heavy quark contributions are given by
\begin{eqnarray}
\label{eqHLg}
H_{L,g}^{\rm S}\left(x, a_s, \frac{Q^2}{m^2}\right) &=&  a_s 
\widehat{c}_{L,g}^{(1)}
+ a_s^2 \left[\frac{1}{2} \widehat{P}_{qg}^{(0)} c_{L,q}^{(1)} \ln 
\left(\frac{Q^2}{m^2}\right) + \widehat{c}_{L,g}^{(2)} \right]\nonumber\\
&+& a_s^3 \Biggl\{\Biggl[\frac{1}{8} \widehat{P}_{qg}^{(0)}
\left[P_{qq}^{(0)}-P_{gg}^{(0)} + 2\beta_0 \right] 
\ln^2\left(\frac{Q^2}{m^2}\right)
+ \frac{1}{2} \widehat{P}^{(1)}_{qg} \ln\left(\frac{Q^2}{m^2}\right)
\nonumber\\
& &~~~~~+ a_{Qg}^{(2)} + \overline{a}_{Qg}^{(1)} \left[
P_{qq}^{(0)} - P_{gg}^{(0)} + 2 \beta_0\right] \Biggr] c_{L,q}^{(1)}
+ \frac{1}{2} \widehat{P}_{qg}^{(0)} \ln\left(\frac{Q^2}{m^2}\right) 
c_{L,q}^{(2)} + \widehat{c}_{L,g}^{(3)}\Biggr\}
\\
\label{eqHLq}
H_{L,q}^{\rm PS} \left(x, a_s, \frac{Q^2}{m^2}\right) &=& 
a_s^2 \widehat{c}_{L,q}^{{\rm PS},(2)} 
+ a_s^3\left\{\left[ -\frac{1}{8} \widehat{P}_{qg}^{(0)} P_{gq}^{(0)} 
\ln^2\left(
\frac{Q^2}{m^2}\right) 
\right. \right. \nonumber\\ & & \left. \left.
+\frac{1}{2} \widehat{P}_{qq}^{{\rm PS},(1)} 
\ln\left(
\frac{Q^2}{m^2}\right) + a_{Qq}^{{\rm PS},(2)} - \overline{a}_{Qg}^{(1)} P_{gq}^{(0)}\right] c_{L,q}^{(1)}
+ \widehat{c}_{L,q}^{{\rm PS},(3)}\right\}  
\\
H_{L,q}^{\rm NS}\left(x, a_s, \frac{Q^2}{m^2}\right) &=& 
a_s^2\left[ - \beta_{0,Q} c_{L,q}^{(1)} 
\ln\left(\frac{Q^2}{m^2}\right)
+ \widehat{c}_{L,q}^{{\rm NS}, (2)} \right]\nonumber\\
&+& a_s^3 \Biggl\{\left[-\frac{1}{4} \beta_{0,Q} P_{qq}^{(0)} 
\ln^2\left(\frac{Q^2}{m^2}\right) - \frac{1}{2} \widehat{P}_{qq}^{{\rm 
NS},(1)}
\ln \left(\frac{Q^2}{m^2}\right) + a_{qq,Q}^{{\rm NS},(2)} 
+ \frac{1}{4} \beta_{0,Q} \zeta_2 P_{qq}^{(0)}\right] 
\nonumber \\ & & \times c_{L,q}^{(1)} + \widehat{c}_{L,q}^{{\rm NS}, 
(3)}\Biggr\}~.
\end{eqnarray}
The Wilson coefficients for heavy quark production consist of 
terms $\propto (m^2/Q^2)^k, k > 0, k~\epsilon~{\bf N}$ and 
the logarithmic and constant contributions
$\propto \ln^l(Q^2/m^2),~l \geq 0$ for on--shell massive quarks. The 
latter terms do not vanish in the 
limit $m^2 \rightarrow 0$ and can be calculated solving the 
renormalization group equations for $F_i^{Q \overline{Q}}(x,Q^2)$. 
\section{\boldmath The Small-$x$ Limit}
\label{sec:SX}

\vspace{1mm}\noindent
In the small $x$ limit the heavy quark Wilson  coefficient at $O(a_s)$, 
$H_{L,g}^{(1)}(x,Q^2/m^2;\mu^2/m^2)$,  vanishes with $x$ 
since its leading pole is situated at $N = -1$. One expects the 
following leading 
and next-to-leading small-$x$ behaviour 
\begin{eqnarray}
H_{L}^{S}(x) \propto~a_s^2~\frac{d_1^{(1)}}{x} + \sum_{k=2}^\infty 
a_s^{k+1} \left[
d_k^{(1)} \frac{\ln^{k-1}(x)}{x} + 
d_k^{(2)} \frac{\ln^{k-2}(x)}{x} + \ldots \right]~. 
\end{eqnarray}
In $O(a_s^2)$ the leading \cite{CCH,BUZA1}
small--$x$ terms for $\mu^2 = Q^2$ are 
\begin{eqnarray}
d_{1,i}^{(1)} &=& - 32 C_i T_R \frac{1}{9}~,
\end{eqnarray}
with $i = A,F$ for the gluonic and pure singlet contribution, 
respectively. As seen in Eqs.~(\ref{eq1a},
\ref{eq1b},
\ref{eq1c},
\ref{eq1d},
\ref{eq1e}), these terms stem 
from the small 
$x$ behaviour of $P_{gg}^{(0)}$ and $\widehat{P}_{gq}^{(0)}$, 
(\ref{eqPgg0},\ref{eqPgq0}), and from $c_{L,g}^{(2)}, c_{L,q}^{(2){\rm PS}}$. 
The terms $\propto T_R$ for both contributions scale by the 
color factors, $C_A,$ resp. $C_F$. $H_{L,g}^{\rm S}$ contains an additional term at $O(a_s^3)$, $\propto 
T_R^2 \ln(Q^2/m^2)$. 
$H_{L,q}^{(2),{\rm NS}}$ does not contain a term 
$\propto 1/x$ but is less singular for small values of $x$.
The corresponding coefficients are obtained in finding the contributions $\propto 1/(N-1)$
in (\ref{eqHLg},\ref{eqHLq}). Most of the respective functions were given above. We further note that
$c_{L,q}^{(1)}(N=1) = 2 C_F$ and $c_{L,q}^{(2)}(N \rightarrow 1) \propto -(32/9) C_F T_R N_f /(N-1)$,
\cite{ZN,BK1}.

In $O(a_s^3)$ the leading small $x$ contributions to 
$H_{L,g(q)}^{{\rm S}, (3)}(x,Q^2/m^2;\mu^2/m^2)$
result from the small $x$ terms in $\widehat{c}^{(3)}_{L,q(g)}(x)$ only 
and are proportional to that of the light flavor contributions 
\cite{FLNNLO-xb}.
The remaining heavy flavor corrections are less singular. The leading 
terms  $\propto 1/x$ are
\begin{eqnarray}
d_{2, i}^{(1)} &=& \frac{128}{3} C_A C_i T_R \left[ - \frac{34}{9} +  
\zeta_2 \right] \\
d_{2, A}^{(2)} &=& -32 C_A C_F T_R\left[ \frac{1}{3}\ln^2\left(\frac{Q^2}{m^2}\right) 
- \frac{10}{9} \ln\left(\frac{Q^2}{m^2}\right) + 
\frac{28}{27}\right]
\nonumber\\
& & - \frac{256}{27} C_F T_R^2 (2 N_F +1) \ln\left(\frac{Q^2}{m^2} \right) 
\nonumber\\
& &+ \frac{32}{3} C_A^2 T_R \left[- \frac{2756}{27} + \frac{65}{3} \zeta_2 
+ 20 \zeta_3\right] 
+ \frac{64}{3} C_A C_F T_R 
\left[\frac{56}{9} - \zeta_2 - 4 \zeta_3\right] 
\nonumber\\ & &
+ C_F T_R^2 (2 N_F + 1) 
\frac{64}{9}\left[\frac{121}{9} - 4 \zeta_2\right]
+ C_A T_R^2 (2 N_F + 1) 
\frac{32}{9}\left[\frac{101}{9} - 8 \zeta_2\right]
~.
\\
d_{2, F}^{(2)} &=& - 32 C_F^2 T_R \left[
\frac{1}{3}\ln^2\left(\frac{Q^2}{m^2}\right) 
- \frac{10}{9} \ln\left(\frac{Q^2}{m^2} \right)  + 
\frac{28}{27}  \right] 
\nonumber\\
& &+ 32 C_A C_F T_R \left[- \frac{899}{27} + 7 \zeta_2 + \frac{20}{3} 
\zeta_3\right] + \frac{64}{3} C_F^2 T_R \left[\frac{56}{9} - \zeta_2 
- 4 \zeta_3\right] 
\nonumber\\ & &
+ C_F T_R^2 (2 N_F + 1) 
\frac{256}{9}\left[\frac{53}{9} - \zeta_2\right]~.
\end{eqnarray}
Among the subleading terms not stemming from $\hat{c}_{L,g(q)}^{(3)}$
those $\propto T_R$ scale by the color factor.
$H_{L,q}^{{\rm NS}, (3)}(x,Q^2/m^2;\mu^2/m^2)$ is regular for $N=1$.
Similarly to the treatment in \cite{KL,BV1} one might consider its 
singular behaviour around $N=0$, however, the small--$x$ resummation for 
these terms were not yet derived.   
\section{Numerical Results}
\label{sec:NR}

\vspace{1mm}\noindent
In the following we will give some numerical illustrations of the effect 
of the heavy flavor contributions in the limit $Q^2 \gg m^2$. These results 
are, unfortunately, of limited phenomenological use, since one 
expects, similar to the case of the NLO corrections \cite{BUZA1}, that 
these 
corrections become effective at large values of $Q^2 \simeq 1000 \GeV^2$, 
where no data on $F_L$ are available at present. At lower values of $Q^2$ power 
corrections do still contribute. In the case of $F_2(x,Q^2)$, a sufficient 
description by the asymptotic expression could be obtained for scales 
$Q^2 \gsim 30 \GeV^2$ already, cf.~\cite{BUZA1}.
  
We illustrate the size of the contributions at NLO and NNLO for 
$F_L^{Q\overline{Q}}(x,Q^2)$ choosing the parton distributions as follows 
at $Q^2_0 = 30 \GeV^2$
\begin{eqnarray}
\label{eqqns}
xq_{\rm NS}(x) &=& N_{\rm NS} x^{a_{\rm NS}} (1-x)^{b_{\rm NS}}
\\
\label{eqqs}
xq_{\rm PS}(x) &=& N_s x^{a_s}\left[(1-x)^{b_s} + c_s 
x^{d_s}\right] 
\\
\label{eqgl}
xg(x) &=& N_g x^{a_g}\left[(1-x)^{b_g} + c_g x^{d_g}\right] 
\end{eqnarray}
We apply this parameterization both for the NLO and NNLO effects for illustrative purposes.
The parton distributions at higher scales are obtained by evolution. The values of the parameters
at $Q_0^2$ are listed in Table~1.
The strong coupling constant $\alpha_s(Q^2)$ is calculated using the values of 
$\Lambda_{\rm QCD}^{\rm 
\overline{MS},~NS,(4)}$ determined in \cite{BBG}. One obtains:
$\alpha_s^{\rm NLO}(30 \GeV^2) = 0.1977,$ resp.~$0.1708 (Q^2 = 100 \GeV^2),~0.1132 (Q^2 = 10^4 \GeV^2)$
and
$\alpha_s^{\rm NNLO}(30 \GeV^2) = 0.1928,~ 0.1673 (Q^2 = 100 \GeV^2),~0.1118 (Q^2 = 10^4 \GeV^2)$.
The charm quark mass was chosen to be $m_c = 1.5 \GeV$.
%
\renewcommand{\arraystretch}{1.3}
%
\begin{center}
\begin{tabular}{||l|r|r|r|r|r||}
\hline \hline
Parameter   & $N_i$      & $a_i$        & $b_i$         & $c_i$  &  $d_i$  \\
\hline \hline
{\rm NS}    & 1.00000    & 0.50000      & 3.0000        &        &  \\
{\rm S}     & 0.60000    &-0.30000      & 3.5000        & 5.0000 & 0.80000 \\
gluon       & 0.11518    &-0.32230      & 6.0445        & 0.9618 & 0.00422 \\
\hline \hline
\end{tabular}
\end{center}
\vspace{2mm}
\noindent
\normalsize
\begin{center}
{\sf Table~1: The parameters 
of the quark non-singlet, singlet and gluon distribution at $Q^2 = 30 \GeV^2$.
}
\end{center}
\renewcommand{\arraystretch}{1.0}

\vspace{3mm}\noindent
\normalsize

The asymptotic heavy flavor non--singlet contributions together with the light flavor terms are shown in 
Figure~1 and turn out to be small due to the
shape of the input distribution and since they emerge only at $O(a_s^2)$. Their contribution shrinks with growing
$Q^2$. The asymptotic heavy flavor pure singlet part added to the light-flavor contributions of $F_L(x,Q^2)$  
are depicted in Figure~2. Also here the the first contribution is 
obtained at 
$O(a_s^2)$, but the effect is  
much larger if compared to the non--singlet part due to the small--$x$ behaviour of the corresponding distribution function 
$\propto 
x^{-0.3}$. The contribution rises with $Q^2$ and amounts to $\sim 
O(1/5 ...1/6)$ of the gluon contributions, 
shown in Figure~3.
The gluon contribution is the largest and   
emerges already at LO. It grows towards the small $x$ region.
In all cases the NNLO result is larger than that at NLO. To obtain a complete picture for the
heavy flavor contributions to $F_L(x,Q^2)$ the corrections at scales of lower values of $Q^2$ have to be calculated. 
\vspace*{-1.2cm}
\begin{center}

\mbox{\epsfig{file=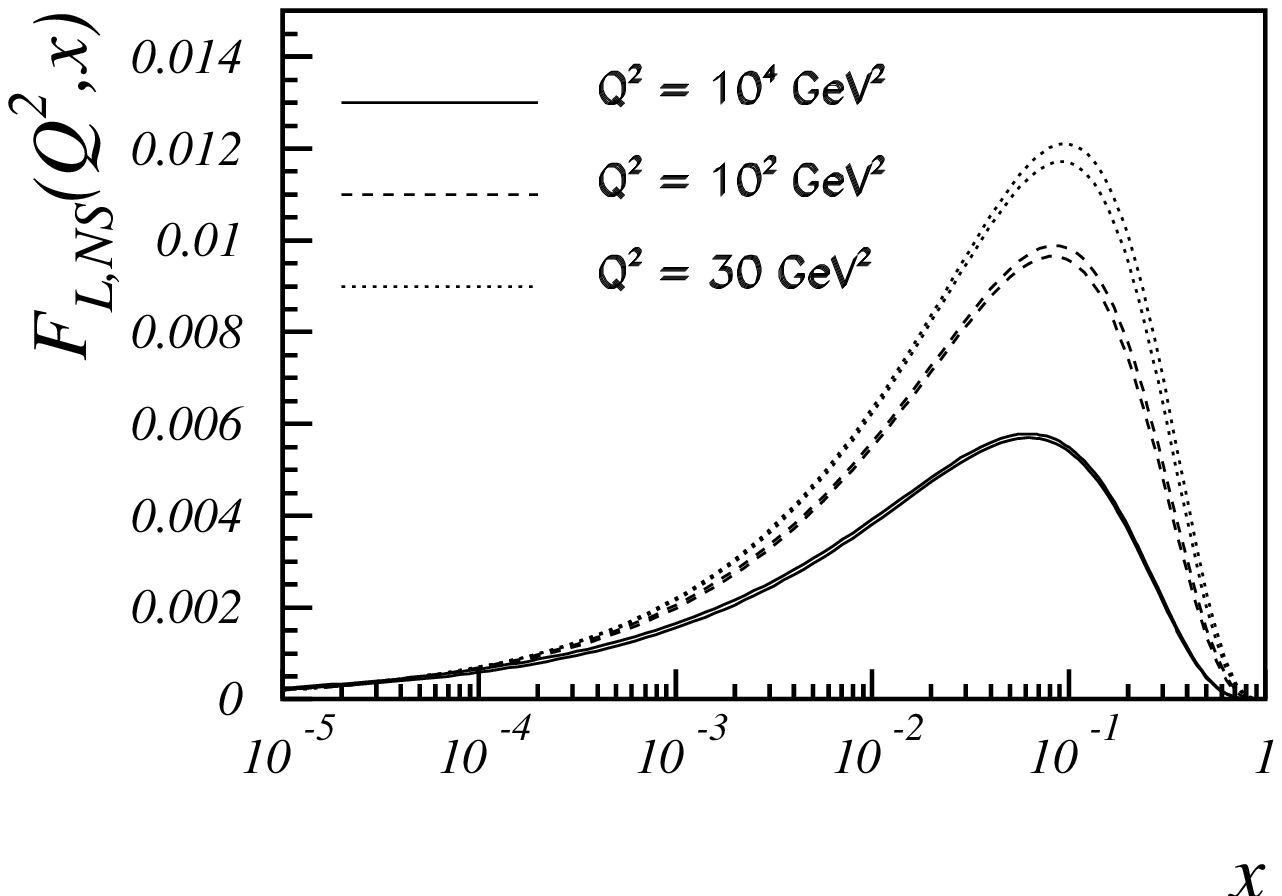,width=15cm}}

\vspace{2mm}
\noindent     
\small   
\end{center}
{\sf
Figure~1:~
The light flavor and asymptotic heavy flavor non-singlet contributions due to charm to 
$F_L(x,Q^2)$ in NLO and NNLO.  
Upper lines: NNLO, lower lines NLO.} 
\normalsize

\vspace{3mm}\noindent

\vspace*{-1cm}
\begin{center}

\mbox{\epsfig{file=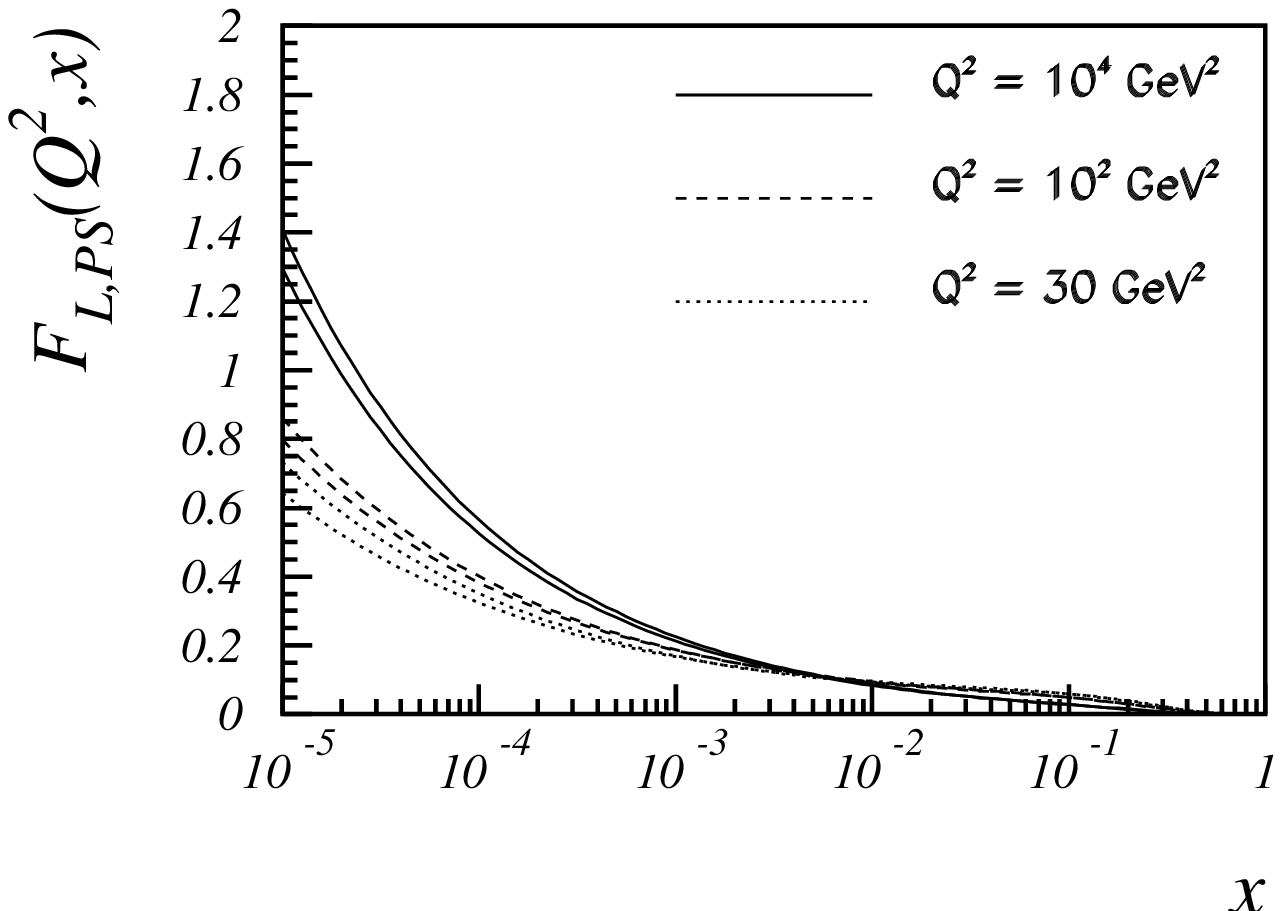,width=15cm}}

\vspace{2mm}
\noindent     
\small   
\end{center}
{\sf
Figure~2:~
The light flavor and asymptotic heavy flavor pure  singlet contributions due to charm to 
$F_L(x,Q^2)$ in NLO and NNLO.  
Upper lines: NNLO, lower lines NLO.} 
\normalsize

\vspace*{-1cm}
\begin{center}

\mbox{\epsfig{file=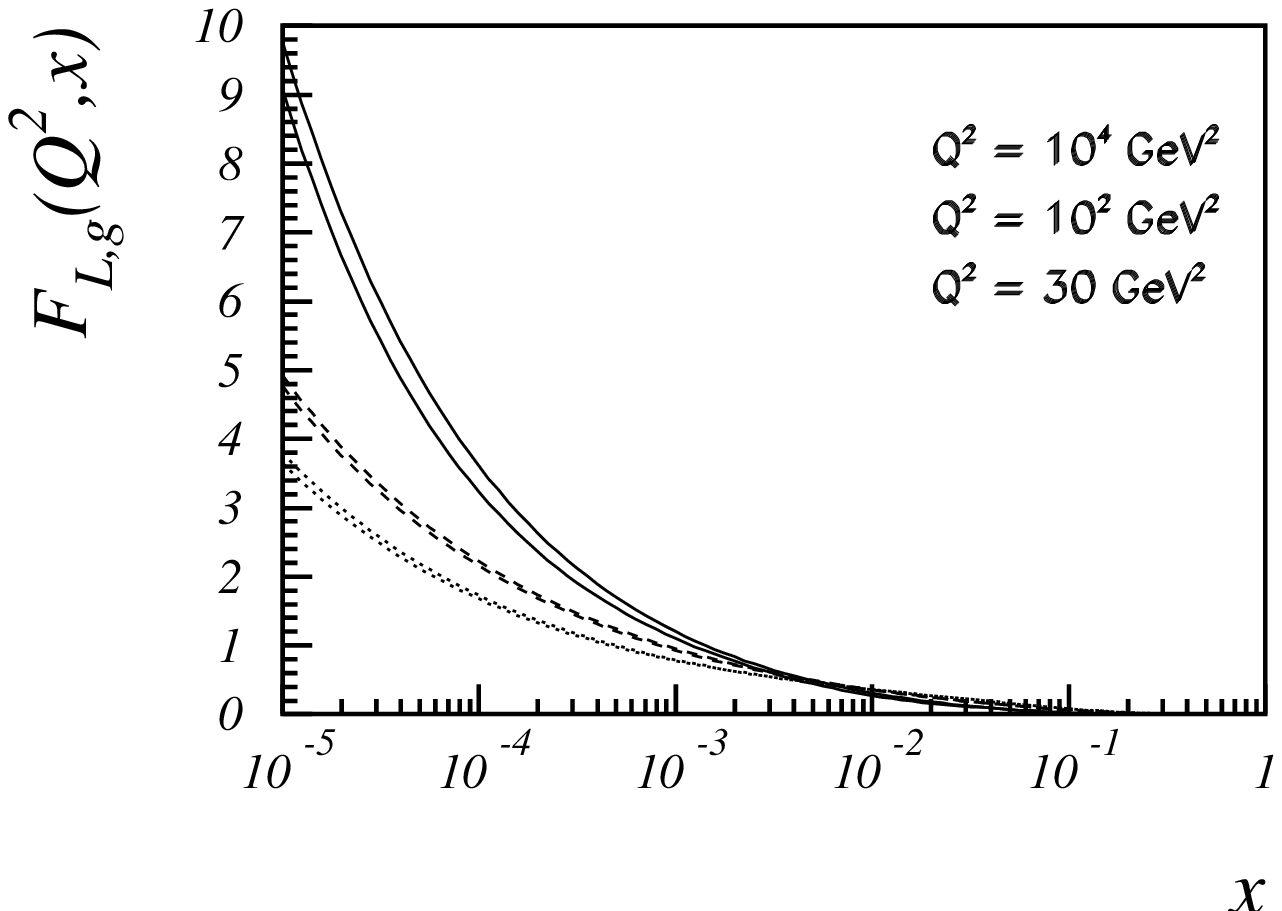,width=15cm}}

\vspace{2mm}
\noindent     
\small   
\end{center}
{\sf
Figure~3:~
The light flavor and asymptotic heavy flavor gluon contributions due to charm to 
$F_L(x,Q^2)$ in NLO and NNLO. 
Upper lines: NNLO, lower lines NLO.} 
\normalsize

\section{Conclusions}
\label{sec:Conc}

\vspace{1mm}\noindent
We have calculated the heavy flavor contributions to the structure function
$F_L(x,Q^2)$ in the region $Q^2 \gg m_Q^2$ at $O(\alpha_s^3)$. In this
kinematic regime the respective terms are obtained as the logarithmic
orders $\propto \ln^k(Q^2/m_Q^2)$ and the constant term. Power corrections 
cannot be determined using the method of the present paper. 
In this approximation the heavy flavor Wilson coefficients are given by
a convolution of the light--flavor Wilson coefficients and {\sf universal} 
operator matrix elements, which contain the information on the heavy quarks. 
At NLO a numerical comparison of the complete calculation 
to the asymptotic case was possible and scales in the range $Q^2 \gsim 1000
\GeV^2$ were identified to apply the asymptotic relation for $F_L^{Q\overline{Q}}(x,Q^2)$. 
This is likely to be the case at  NNLO too. We presented numerical results for the asymptotic
$O(\alpha_s^3)$ corrections added to the light--flavor contributions.
The largest contribution is due to the gluonic term, followed by the pure singlet term, which is
a factor $\sim 5$ smaller in the small $x$ region. The flavor non--singlet contribution
is very small. The leading small $x$ terms of the heavy flavor Wilson coefficients $H_{L, q(g)}^{\rm 
PS,(S)}(x, Q^2/m^2)$
were determined. In $O(a_s^3)$ the pure heavy flavor terms contribute to the next-to-leading
small $x$ terms. In part of the terms scaling by $C_F (C_A)$ is observed comparing the respective gluonic and 
quarkonic contributions.
\section{Appendix}
\label{sec:appendix}

\vspace{1mm}\noindent
The Mellin transforms used in the present calculation may be found in 
\cite{BK1}. Some expressions can be written in the more compact form  
given below.
\begin{eqnarray}
\Mvec[\ln(1+z)](N) &=& \frac{1}{N}\left\{\ln(2)- (-1)^N 
\left[S_{-1}(N)+\ln(2)\right]\right\} 
\nonumber\\ &=& \frac{1}{N} \left[\ln(2) - \beta(N+1) \right]\\
\Mvec[\ln(z)\ln(1+z)](N) &=& -\frac{1}{N^2} \left[\ln(2) - \beta(N+1) 
\right] - \frac{1}{N} \beta'(N+1)\\
\Mvec[\ln^2(z) \ln(1+z)](N) &=& \frac{2}{N^3}\left\{\ln(2) - (-1)^N 
\left[S_{-1}(N)+\ln(2)\right]\right\} \nonumber\\ & &- (-1)^N 
\frac{2}{N^2} 
\left[S_{-2}(N)+\frac{\zeta_2}{2}\right] 
- (-1)^N \frac{2}{N}\left[S_{-3}(N) + \frac{3}{4} \zeta_3\right]
\nonumber\\
&=& \frac{2}{N^3} \left[ \ln(2) - \beta(N+1) \right]
+\frac{2}{N^2} \beta'(N+1) - \frac{1}{N} \beta''(N+1)\nonumber\\
\\
\Mvec[\Li_2(-z)](N) &=& - \frac{\zeta_2}{2N}+\frac{1}{N^2}\left\{
\ln(2)-(-1)^N\left[S_{-1}(N)+\ln(2)\right]\right\} \nonumber\\
&=& - \frac{\zeta_2}{2N}+\frac{1}{N^2}\left[
\ln(2)-\beta(N+1)\right] \\
\Mvec[\ln(z) \Li_2(-z)](N) &=& \frac{\zeta_2}{2N^2}-\frac{2}{N^3}\left[
\ln(2)-\beta(N+1)\right] - \frac{1}{N^2} \beta'(N+1)\\
\Mvec[\Li_2(-z) + \ln(z) \ln(1+z)](N) &=& 
- \frac{1}{2 N}\left[\zeta_2 + 2 \beta'(N+1)\right]\\
\Mvec[\Li_3(-z)](N) &=& - \frac{3}{4N} \zeta_3+ 
\frac{1}{2 N^2} \zeta_2  - \frac{1}{N^3} \left[\ln(2) - \beta(N+1) \right]
\\
\Mvec[\Phi_1(z)](N) &=& \frac{(-1)^{N+1}}{N}\left\{2 S_{1,-2}(N)
+ \zeta_2 \left[S_1(N) - S_{-1}(N)\right]\right\}
\nonumber\\
& & +\frac{\left[1+(-1)^{N+1}\right]}{N} \left[\frac{\zeta_3}{4} - \zeta_2 
\ln(2)\right]\\
&=& \frac{1}{N} \left\{2 \Mvec\left[\frac{\Li_2(x)}{1+x}\right](N)
- \frac{2}{N} \zeta_2 + \frac{2}{N^2} S_1(N)
    + 3 \zeta_2 \beta(N+1) 
\right. \nonumber\\ & &
+ 2 S_1(N) \beta'(N+1) 
- \beta''(N+1)  
\left. + \frac{\zeta_3}{4} - \zeta_2 \ln(2) 
\right\}~,
\end{eqnarray}
where
\begin{eqnarray}
\Phi_1(z) = 2 \Li_2(-z) \ln(1+z) + \ln^2(1+z) \ln(z) + 2 S_{1,2}(-z)~.
\end{eqnarray}
$\Mvec\left[\Li_2(x)/(1+x)\right](N)$ is a basic function, 
cf.~\cite{ANCONT,ALGEBRA}.

\vspace{3mm}
\noindent
{\bf Acknowledgment}.~
We are indebted for discussions to H. B\"ottcher, E. Christy, R.~Ent, 
A.~Guffanti, and S. Moch. This work was supported in part by DFG 
Sonderforschungsbereich Transregio 9, Computergest\"utzte 
Theoretische Physik.
\newpage


\begin{thebibliography}{999}
%
\bibitem{PDG}
Particle Data Group, S. Eidelman et al., Phys. Lett. {\bf B592} (2004) 1.
%
\bibitem{FTARG}
L.W. Whitlow et al. Phys. Lett. {\bf B250} (1990) 193;\\
S. Dasu et al., E140 collab., Phys. Rev. {\bf D49} (1994) 5641;\\
L.H. Tao et al.,  E140x collab., Z. Phys. {\bf C70} (1996) 387;\\
M. Arneodo et al., NMC collab., Nucl. Phys. {\bf B483} (1997) 
3; {\bf B487} (1997) 3;\\
Y. Liang et al. {\tt  nucl-ex/0410027}.
%
\bibitem{H1}
C. Adloff et al., H1 collab., Phys. Lett. {\bf B393} (1997) 452.
%
\bibitem{MK}
M. Klein, in~: Proc of the 12th Int. Workshop on Deep Inelastic 
Scattering, DIS 2004, Strebske Pleso, 2004, pp.~309;\\
J. Feltesse, {\sf On a measurement of the longitudinal structure function
$F_L$ at HERA}, Proc. of the 2005 Ringberg Workshop on {\sf New Trends in 
HERA Physics, 2005}, eds. B. Kniehl, G.~Kramer, and W. Ochs;\\
M.~Dittmar {\it et al.},
arXiv:hep-ph/0511119.
%
\bibitem{CG}
C.G. Callan, jr. and D.J. Gross, Phys. Rev. Lett. {\bf 22} (1969) 156. 
%
\bibitem{GP}
H. Georgi and H.D. Politzer, Phys. Rev. {\bf D14} (1976) 1829.
%
\bibitem{FLLO}
A. Zee, F. Wilczek, and S.B. Treiman, Phys. Rev. {\bf D10} (1974) 2881.
%
\bibitem{FLNLO}
D.W. Duke, J.D. Kimel, and G.A. Sowell, Phys. Rev. {\bf D25} (1982) 71;\\
A.~Devoto, D.W. Duke, J.D. Kimel, and G.A. Sowell, Phys. Rev. {\bf D30} 
(1984) 541;\\
D.I. Kazakov and A.V. Kotikov, Nucl. Phys. {\bf B307} (1988) 721; Phys. 
Lett. {\bf
B291}  (1992) 171;\\
D.I. Kazakov, A.V. Kotikov, G.~Parente, O.A.~Sampayo, and J.~Sanchez
Guillen Phys. Rev. Lett. {\bf 65} (1990) 1535;\\
J.~Sanchez Guillen, J.~Miramontes, M.~Miramontes, G.~Parente, and
O.A. Sampayo, Nucl. Phys. {\bf B353} (1991) 337;\\
S.A. Larin, J.A.M. Vermaseren, Z. Phys. {\bf C57} (1993) 93.
%
\bibitem{ZN}
W.L. van Neerven and E.B. Zijlstra, Phys. Lett. {\bf B272} (1991) 127;\\
E.B. Zijlstra and W.L. van Neerven, Phys. Lett. {\bf B273} (1991) 476; 
Nucl. Phys.
{\bf B383} (1992) 525.
%
\bibitem{MV}
S. Moch and J.A.M.~Vermaseren, Nucl. Phys. {\bf B573} (2000) 853.
%
\bibitem{FLNNLO-mom}
S.A. Larin, T.~van Ritbergen, and J.A.M. Vermaseren,
Nucl. Phys. {\bf B427} (1994) 41;\\
S.A. Larin, P.~Nogueira, T.~van Ritbergen, and J.A.M. Vermaseren,
Nucl. Phys. {\bf B492} (1997) 338;\\
A. Retey and J.A.M. Vermaseren, Nucl. Phys. {\bf B604} (2001) 281;\\
J. Bl\"umlein and J.A.M. Vermaseren, Phys. Lett. {\bf B606} (2005) 130.
%
\bibitem{FLNNLO-xa}
S.-O. Moch, J.A.M. Vermaseren, and A. Vogt, Phys. Lett. {\bf B606} (2005) 
123.
%
\bibitem{FLNNLO-xb}
J.A.M. Vermaseren, A. Vogt, and S.-O. Moch, Nucl. Phys. {\bf B724} (2005) 
3.
%
\bibitem{DEVEN}
A.M. Cooper--Sarkar, G. Ingelman, K.R. Long, R.G. Roberts and D.H. Saxon,
Z. Phys. {\bf C39} (1988) 281.
%
\bibitem{CH}
S. Catani and F. Hautmann, Nucl. Phys. {\bf B427} (1994) 475.
%
\bibitem{JBFL}
J. Bl\"umlein J. Phys. {\bf G19} (1993) 1623; Nucl. Phys. (Proc. Suppl.) 
{\bf 39BC} (1995) 22.
%
\bibitem{BV}
J. Bl\"umlein and A. Vogt, Phys. Rev. {\bf D57} (1998) R1; {\bf D58} 
(1998) 014020.
%
\bibitem{HTW}
R.K. Ellis, W. Furmanski, and R. Petronzio, Nucl. Phys. {\bf B212} (1983) 
29;\\
J.-W. Qiu, Phys. Rev. {\bf D42} (1990) 30;\\
J. Bartels, C. Bontus, and H. Spiesberger, {\tt hep-ph/9908411}; \\
J. Bartels and C. Bontus, Phys. Rev. {\bf D61} (2000) 034009.
%
\bibitem{BR}
J. Bl\"umlein  and S. Riemersma, {\tt hep-ph/9609394}.
%
\bibitem{FLQQLO}
E. Witten, Nucl. Phys. {\bf B104} (1976) 445;\\
J. Babcock and D. Sivers, Phys. Rev. {\bf D18} (1978) 2301;\\
M.A. Shifman, A.I. Vainshtein, and V.I. Zakharov, Nucl. Phys. {\bf B136} 
(1978) 157;\\
J.P. Leveille and T. Weiler, Nucl. Phys. {\bf B147} (1979) 147;\\
M. Gl\"uck, E. Hoffmann, and E. Reya, Z. Phys. {\bf C13} (1982) 119.
%
\bibitem{FLQQNLO1}
E. Laenen, S. Riemersma, J. Smith, and W.L. van Neerven, Nucl. Phys. {\bf 
B392} (1993) 162; 229.
%
\bibitem{FLQQNLO2}
S. Riemersma, J. Smith, and W.L. van Neerven, 
Phys. Lett. {\bf B347} (1995) 143.
%
\bibitem{SAJB}
S.I. Alekhin and J. Bl\"umlein, Phys. Lett. {\bf B594} (2004) 299.
%
\bibitem{BUZA1}
M. Buza, Y. Matiounine, J. Smith, R.L. Migneron and W.L. van Neerven,
Nucl. Phys. {\bf B472} (1996) 611.
%
\bibitem{BUZA2}
M. Buza, Y. Matiounine, J. Smith,  and W.L. van Neerven, Nucl. Phys. {\bf 
B485} (1997) 420.
%
\bibitem{BUZA3}
M. Buza and W.L. van Neerven, Nucl. Phys. {\bf 500} (1997) 301;\\
M. Buza, W.L. van Neerven, and J. Smith, Eur. Phys. J. {\bf C1} (1998) 
301.
%
\bibitem{BK1}
J.~Bl\"umlein and S.~Kurth,
Phys.\ Rev.\ D {\bf 60} (1999) 014018
[arXiv:hep-ph/9810241];
{\tt arXiv:hep-ph/9708388.}
%
\bibitem{ALGEBRA}
J. Bl\"umlein, Comp. Phys. Commun. {\bf 159} (2004) 19; Few Body Syst. 
{\bf 36} (2005) 29.
%
\bibitem{JBVR}
J. Bl\"umlein and V. Ravindran,
Nucl. Phys. {\bf B716} (2005)128; {\tt hep-ph/0604019}, Nucl. Phys. {\bf B} (2006) in print.
%
\bibitem{ANCONT}
J. Bl\"umlein, Comput. Phys. Commun. {\bf 133} (2000) 76. 
%
\bibitem{JBSM}
J. Bl\"umlein and S.-O. Moch, Phys. Lett. {\bf B614} (2005) 53.
%
\bibitem{JB05}
J. Bl\"umlein, in preparation.
%
\bibitem{CCH}
S. Catani, M. Ciafaloni,  and F. Hautmann, Nucl. Phys. {\bf B366} (1991) 
135.
%
\bibitem{KL}
R. Kirschner and L. Lipatov, Nucl. Phys. {\bf B213} (1983) 122.
%
\bibitem{BV1}
J. Bl\"umlein and A. Vogt, Phys. Lett. {\bf B381} (1996) 296; Acta Phys. 
Pol. {\bf B27} (1996) 1309. 
%
\bibitem{BBG}
J.~Bl\"umlein, H.~B\"ottcher and A.~Guffanti,
arXiv:hep-ph/0607200.
\end{thebibliography}
\end{document}